\documentclass[]{spie}

\usepackage[]{graphicx}
\usepackage[]{graphicx}

\usepackage[numbers]{natbib}

\def\deg{\hbox{$^\circ$}}

\def\lae{\mathrel{\raise .4ex\hbox{\rlap{$<$}\lower 1.2ex\hbox{$\sim$}}}}
\def\gae{\mathrel{\raise .4ex\hbox{\rlap{$>$}\lower 1.2ex\hbox{$\sim$}}}}

\def\apj{{ApJ}}

\def\procspie{{Proceedings SPIE}}

\title{Updating the Chandra HETGS Efficiencies using In-Orbit Observations} 

\author{Herman L.\ Marshall\supit{a}
\skiplinehalf
\supit{a}MIT Kavli Institute, Cambridge, MA, USA
}

\authorinfo{Further author information: (Send correspondence to H.L.M.)\\H.L.M.: hermanm@space.mit.edu,
Telephone: 1 617 253 8573} 

\begin{document} 
  \maketitle 

\begin{abstract}
The efficiencies of the gratings in the High Energy Transmission Grating Spectrometer (HETGS)
were updated using in-flight observations of bright continuum sources.
The procedure first involved verifying that fluxes obtained from the $+1$ and $-1$
orders match, which checks that the contaminant model and the CCD quantum efficiencies agree.
Then the fluxes derived using the high energy gratings (HEGs) were compared to
those derived from the medium energy gratings (MEGs).
The flux ratio was fit to a low order polynomial, which was allocated to the MEGs above 1 keV or the HEGs below 1 keV.
The resultant efficiencies were tested by examining fits to blazar spectra.
\end{abstract}



\section{Introduction}

This is an update to the
{\em Chandra} High Energy Transmission Grating (HETG) calibration based
on in-orbit observations.
The HETG was described by Canizares et al.\ (2005 \cite{hetg})
and previous flight calibration results were reported
by Marshall et al.\ (2004 \cite{2004SPIE.5165..457M}).
There are two grating types on the HETG: the high energy gratings (HEGs) have
$\times$2 higher dispersion than the
medium energy gratings (MEGs).  When combined with the detectors,
the system is called the HETG Spectrometer (HETGS).
 
A new set of HETG efficiencies was released as part of {\it Chandra} CalDB v4.4.7,
which was intended to reduce systematic differences between the HEG and
MEG fluxes as well as improve overall model fits.  Here is provided details of
the steps that went into the release and what is in store for the next
update.  Updates will also be available at the HETGS calibration web site
{\tt http://space.mit.edu/ASC/calib/hetgcal.html} and are presented
regularly at meetings of the International Astronomical Consortium
for High Energy Calibration (IACHEC) (see 
{\tt http://web.mit.edu/iachec/meetings/index.html}).

\section{Data Handling}

First, data were selected from the TGcat on-line catalog of HETGS
spectra.  See {\tt http://tgcat.mit.edu} to obtain the data used here,
which are given in Table~\ref{tab:observations}.
I started with sources of all sorts but relied most on extragalactic
sources with relatively smooth continua.
A simple type is given: ``BLL'' for the BL Lac
objects (without strong optical emission lines), and ``AGN'' for the
remaining quasars and Seyfert galaxies.

  \begin{table}
    \caption{HETGS Calibration Observations}
    \label{tab:observations}
    \centering
    \begin{tabular}{|lcrcc|}
    \hline
     \hline
   Name & Type & ObsID & Year & Exposure \\
     &  & & & (ks) \\
    \hline
3C 273 & AGN & 459 & 2000.30 & 38.6 \\
PKS 2155-304 & BLL & 337 & 2000.38 & 38.6 \\
ARK 564 & AGN & 863 & 2000.95 & 48.7 \\
PKS 2155-304 & BLL & 1705 & 2001.33 & 25.5 \\
MKN 509 & AGN & 2087 & 2001.65 & 57.9 \\
3C273 & AGN & 2463 & 2001.82 & 26.7 \\
PKS2155-304 & BLL & 1014 & 2002.07 & 26.7 \\
NGC 4593 & AGN & 2089 & 2002.35 & 78.9 \\
IC 4329A & AGN & 2177 & 2002.42 & 59.1 \\
3C273 & AGN & 3573 & 2002.58 & 29.7 \\
3C 120 & AGN & 3015 & 2002.58 & 57.2 \\
PKS2155-304 & BLL & 3167 & 2002.80 & 29.6 \\
3C273 & AGN & 4430 & 2003.70 & 27.2 \\
PKS2155-304 & BLL & 3708 & 2003.77 & 26.6 \\
PKS2155-304 & BLL & 3706 & 2003.77 & 27.7 \\
H 1426+428 & BLL & 3568 & 2003.90 & 99.4 \\
3C273 & AGN & 5169 & 2005.38 & 29.7 \\
PKS2155-304 & BLL & 5173 & 2005.57 & 26.7 \\
3C382 & AGN & 4910 & 2005.62 & 54.2 \\
3C382 & AGN & 6151 & 2005.72 & 63.9 \\
1H1426+428 & BLL & 6088 & 2006.22 & 40.4 \\
3C273 & AGN & 8375 & 2008.22 & 29.6 \\
3C273 & AGN & 9703 & 2008.57 & 29.7 \\
Ark 564 & AGN & 9898 & 2008.77 & 99.5 \\
Ark 564 & AGN & 10575 & 2008.83 & 62.2 \\
NGC 4051 & AGN & 10777 & 2009.00 & 27.4 \\
NGC 4051 & AGN & 10775 & 2009.07 & 30.4 \\
NGC 4051 & AGN & 10801 & 2009.23 & 25.7 \\
Ark 564 & AGN & 9899 & 2009.42 & 84.1 \\
MKN421 & BLL & 13098 & 2011.60 & 14.6 \\
\hline
\end{tabular}

\end{table}

\begin{figure}
   \begin{center}
   \begin{tabular}{c}
   \includegraphics[width=15cm]{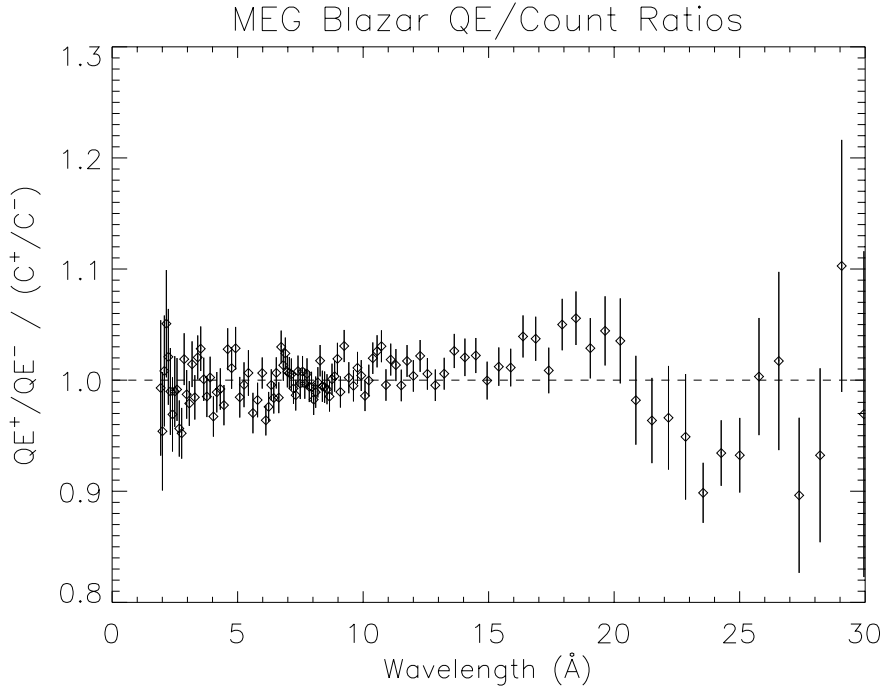}
   \end{tabular}
   \end{center}
\caption{
 Ratio of adaptively binned MEG count rates from the $+1$ order to that of $-1$ order,
 corrected for the ratio of the quantum efficiencies of the detectors.
 The two sides agree to within 10\% everywhere and to better than
 3\% over most energies.
 If the contaminant differed on the two sides, there would be a sharp edge
 at the O-K edge, 23.5 \AA\, which is not observed.
}
\label{fig:rateratio}
\end{figure}

I follow an approach outlined in previous papers on HETGS
effective area \cite{1998SPIE.3444...64M,2004SPIE.5165..457M}.
For the $i$th observation with net counts $C_i^+$ in the $+1$ order and $C_i^-$
net counts in the $-1$ order and each grating, the ratio

\begin{equation}
R = \frac{Q_+}{Q_-} \frac{\sum_i C_i^-}{\sum_i C_i^+}
\end{equation}

\noindent
and its statistical uncertainty $\sigma_R$ were computed over adaptively
sized wavelength bins and a variety of sources.
Note that the plus and minus order efficiencies are assumed to be equal,
which is reasonable from ground tests and the random
$\pm$180\deg\ installation of grating facets in the HETG.
The detector QEs on the $+1$ and $-1$ sides,
$Q_+$ and $Q_-$, are
derived from the models of the CCDs in the detector and depend on
wavelength.
Assuming that the QEs are not perfectly known, $\hat{Q}_+$ and
$\hat{Q}_-$ are assigned to represent the {\em true} QEs on these sides, so that the expected
counts in each wavelength bin on the $+1$ and $-1$ sides are given by

\begin{eqnarray}
\label{eq:counts1}
C_i^+ = n_i A t_i \hat{\epsilon} T \hat{Q}_+ d\lambda \\
C_i^- = n_i A t_i \hat{\epsilon} T \hat{Q}_- d\lambda
\label{eq:counts2}
\end{eqnarray}

\noindent
where $n_i$ is the source flux in
ph cm$^{-2}$ s$^{-1}$ \AA$^{-1}$ for observation $i$, $A$ is the effective area of
the HRMA, $t_i$ is the observation exposure, $\hat{\epsilon}$ is the true
grating efficiency into first order for the grating of interest,
$T$ is the transmission of the detector filter, $Q_+$ ($Q_-$) is
the CCD efficiency on the $+1$ ($-1$) order side, and $d\lambda$ is the
wavelength interval corresponding to one bin.  All quantities are
functions of wavelength except $t_i$.  As defined, the quantity $R$ is

\begin{equation}
\label{eq:meghegratio}
R = \frac{Q_+}{Q_-} \frac{\hat{Q}_-}{\hat{Q}_+}
\end{equation}

\noindent
independent of the source model, grating
efficiency, or filter transmission.
The value of $R$, then, gives the ratio of true QEs relative to the modeled QEs.  If we
have reason to believe that the the $+1$ QEs are correct, for instance,
then $Q_+ = \hat{Q}_+$, so one may obtain the true $-1$ QE via $\hat{Q}_- = R Q_-$
because $R$ is observed and $Q_-$ is the current QE model for the $-1$ side.
See Fig.~\ref{fig:rateratio} for results.
All fluxes came out
to be consistent to within 1\% on average except near 0.5 keV, where the
two sides are consistent at the 3-5\% level but generally within the
uncertainties.

\section{Correcting the MEG/HEG Efficiency Ratio}

With a confirmation that the detector QEs on the $+1$ and $-1$ sides agree for
both HEG and MEG, we now compare MEG to HEG.  First, Eqs.~\ref{eq:counts1}
and \ref{eq:counts2} are recast to combine $+1$ and $-1$ orders:

\begin{eqnarray}
C_i^H = C_i^{-H} + C_i^{+H} = n_i A t_i \hat{\epsilon}^H T (\hat{Q}_{-H} + \hat{Q}_{+H}) d\lambda \\
C_i^M = C_i^{-M} + C_i^{+M} = n_i A t_i \hat{\epsilon}^M T (\hat{Q}_{-M} + \hat{Q}_{+M}) d\lambda
\end{eqnarray}

\noindent
Defining $\hat{Q}_H = \hat{Q}_{-H} + \hat{Q}_{+H}$ and $\hat{Q}_M =
\hat{Q}_{-M} + \hat{Q}_{+M}$, then we form the ratio

\begin{equation}
\label{eq:rmh}
R_{M/H} = \frac{\epsilon_H \hat{Q}_H \sum_i C_i^M}{\epsilon_M \hat{Q}_M \sum_i C_i^H}
\end{equation}

\noindent
where, again, we assume that the grating efficiencies, $\epsilon$, are not
perfectly known so that 

\begin{equation}
\label{eq:pmratio}
R_{M/H} = \frac{\epsilon_H}{\epsilon_M} \frac{\hat{\epsilon}_M}{\hat{\epsilon}_H}
\end{equation}

\noindent
gives the observed quantity that can be used to correct the MEG/HEG grating
efficiency ratio.  Again, as an example, if the HEG efficiencies are assumed
to be correct, giving $\epsilon_H = \hat{\epsilon}_H$, then the observed values
of $R_{M/H}$ can be used to correct the MEG efficiencies by
$\hat{\epsilon}_M = \epsilon_M R_{M/H}$.  Note that $\hat{Q}_H$ does not
cancel $\hat{Q}_M$ in eq.~\ref{eq:rmh} because the ratio is obtained at the same
wavelength for both MEG and HEG but will, in general, be on different detectors
with different QEs due to the dispersion difference.

\begin{figure}
   \begin{center}
   \begin{tabular}{c}
   \includegraphics[width=15cm]{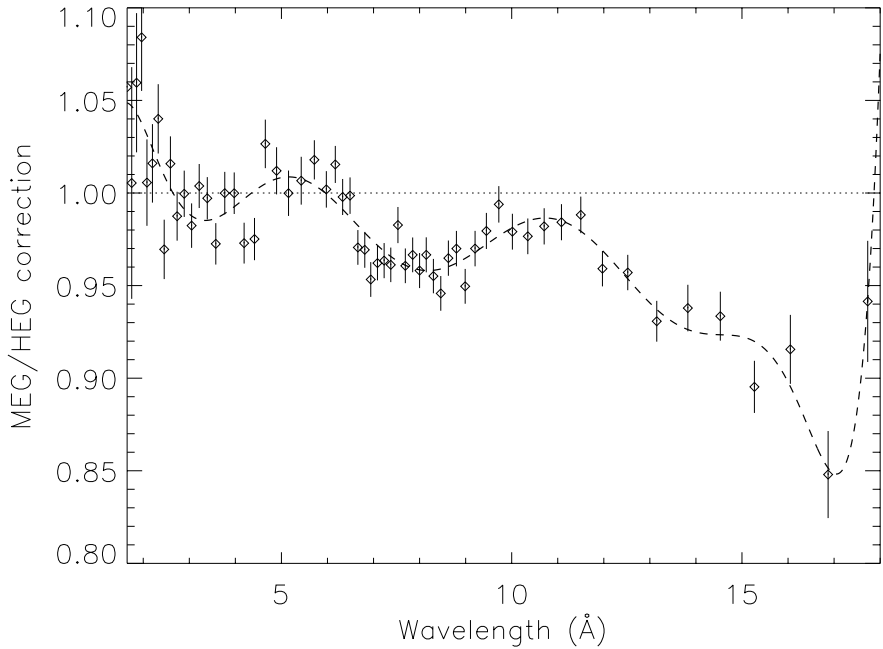}
   \end{tabular}
   \end{center}
\caption{
 The ratio of the MEG and HEG fluxes as a
function of wavelength, binned to have 1\% uncertainties or bin widths
less than 5\% of the central wavelength.
The sense of the correction is that the values shown should be applied
to the MEG efficiencies to bring MEG fluxes into agreement with those
derived from HEG data.  The dashed line is a 9th order polynomial
fit to the data.
}
\label{fig:megheg}
\end{figure}

Results are shown in Fig.~\ref{fig:megheg}.
Values are binned to have 1\% uncertainties so that small
systematic errors may be found and corrected; bins are required to be no
wider than $0.03\lambda$.
A 9th order polynomial was fit to the ratio data to create a correction
factor, $f$, to apply to MEG efficiencies in order to force agreement
between the MEG and HEG.  At this point, one could equally well
apply the correction to just HEG efficiencies, which would require
multiplying them by $1/f$.

\begin{figure}
   \begin{center}
   \begin{tabular}{c}
   \includegraphics[width=15cm]{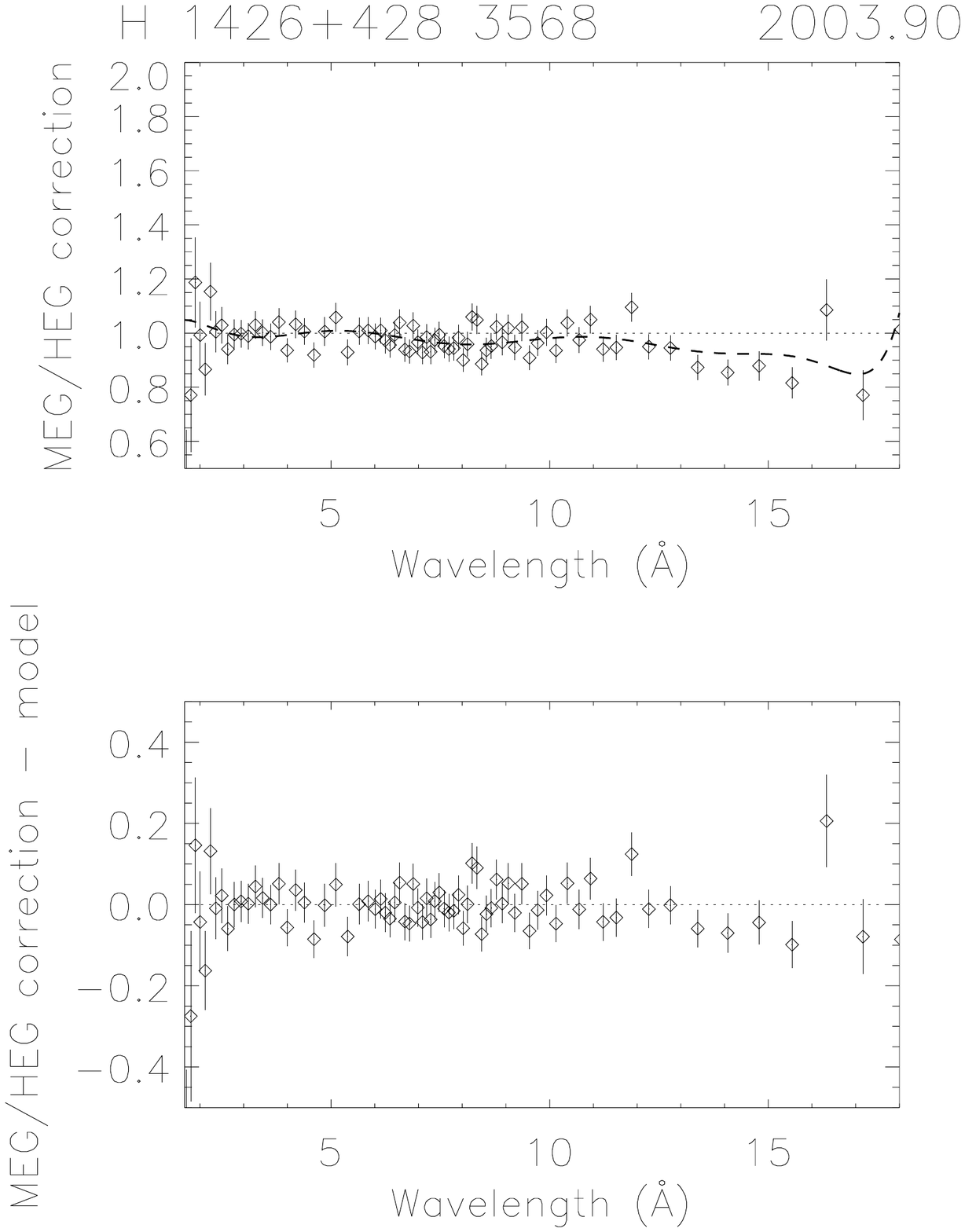}
   \end{tabular}
   \end{center}
\caption{
 Test of the HEG/MEG corrections on one blazar, the BL Lac object 1H 1426$+$428, from obsID 3568,
 observed late in 2003.
 The top panel shows the original ratio data and the spline fit from Fig.~\ref{fig:megheg}.
 The bottom panel shows the residuals after correction, which are consistent with zero.
 }
\label{fig:megheg3568}
\end{figure}

Fig.~\ref{fig:megheg3568} shows the MEG/HEG flux ratio for a specific
observation, obsID 3568 of the BLL source 1H 1426$+$428.
The data clearly follow the trend fit to the average found from combining
all observations; after correction, the residuals are random.  A figure
was generated for each source in order to search for irregularities in
the MEG/HEG ratio but none were found in the data of the other
targets.

\section{Allocating Efficiency Adjustments}

Because the polynomial fit can correct only the ratio of the MEG and
HEG efficiencies, the
correction must be allocated to either the HEG or MEG efficiencies.
The approach chosen was to pick a cross-over wavelength, $\lambda_x$;
for $\lambda > \lambda_x$,
all corrections are applied to the HEG, while for $\lambda < \lambda_x$,
the MEG efficiencies are corrected.  The reasoning behind
the choice is that it was somewhat easier to determine HEG efficiencies
on the ground at high energies due to the higher resolution while harder
to measure them at low energies due to significantly lower effective area.
Furthermore, the HEG part of the HETGS dominates observations at
high energy while the MEG part dominates at low energy, so a ``least harm''
dictum suggests applying less correction where a grating part dominates.
Three values of $\lambda_x$ were tried: 5.3 \AA\ (2.3 keV), 10.8 \AA\ (1.05 keV), and
17.7 \AA\ (0.7 keV).  The last choice effectively assigned the entire
correction to the MEG.

\section{Spectral Fit Results}

\begin{figure}
   \begin{center}
   \begin{tabular}{c}
   \includegraphics[width=13cm]{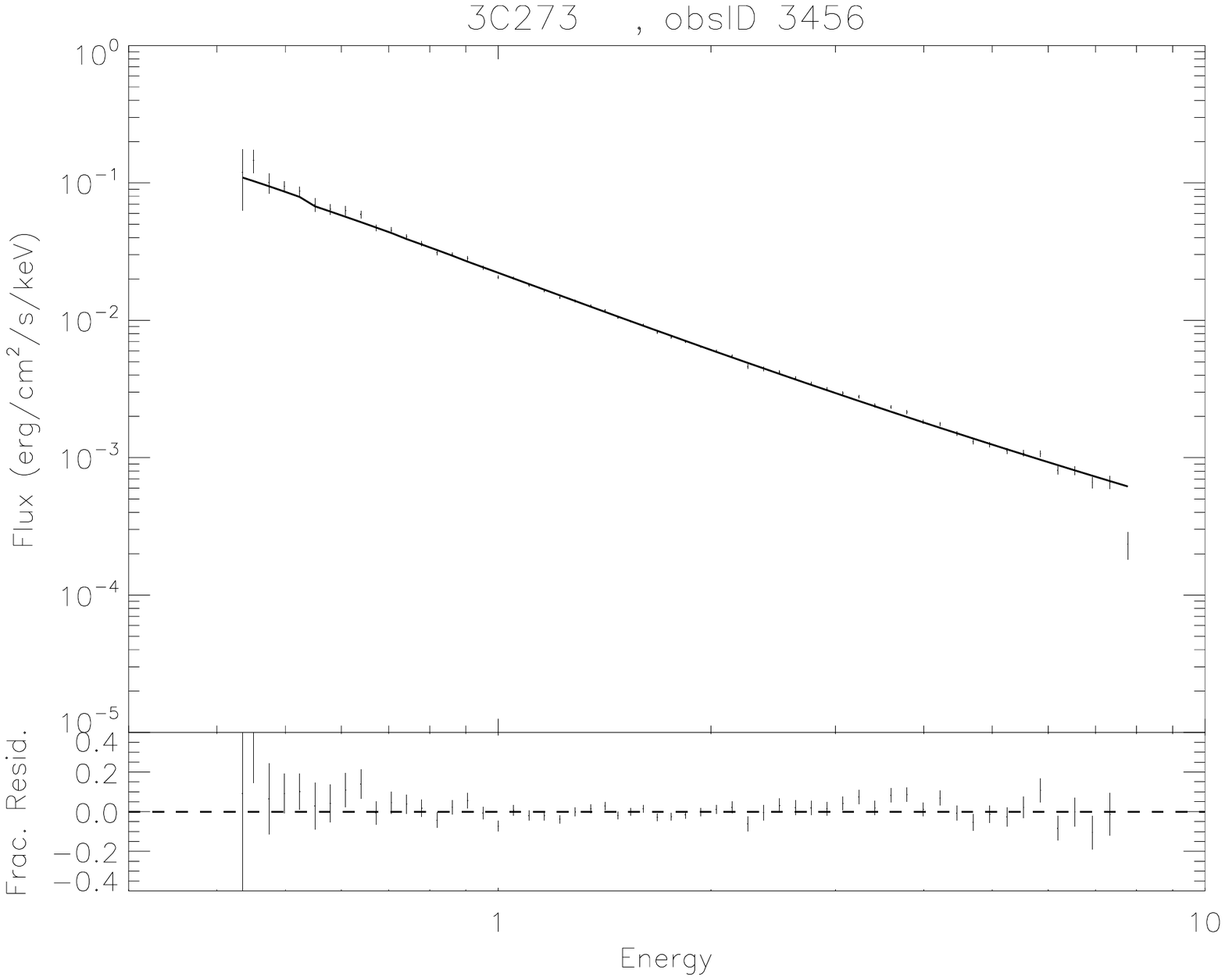} \\
   \includegraphics[width=13cm]{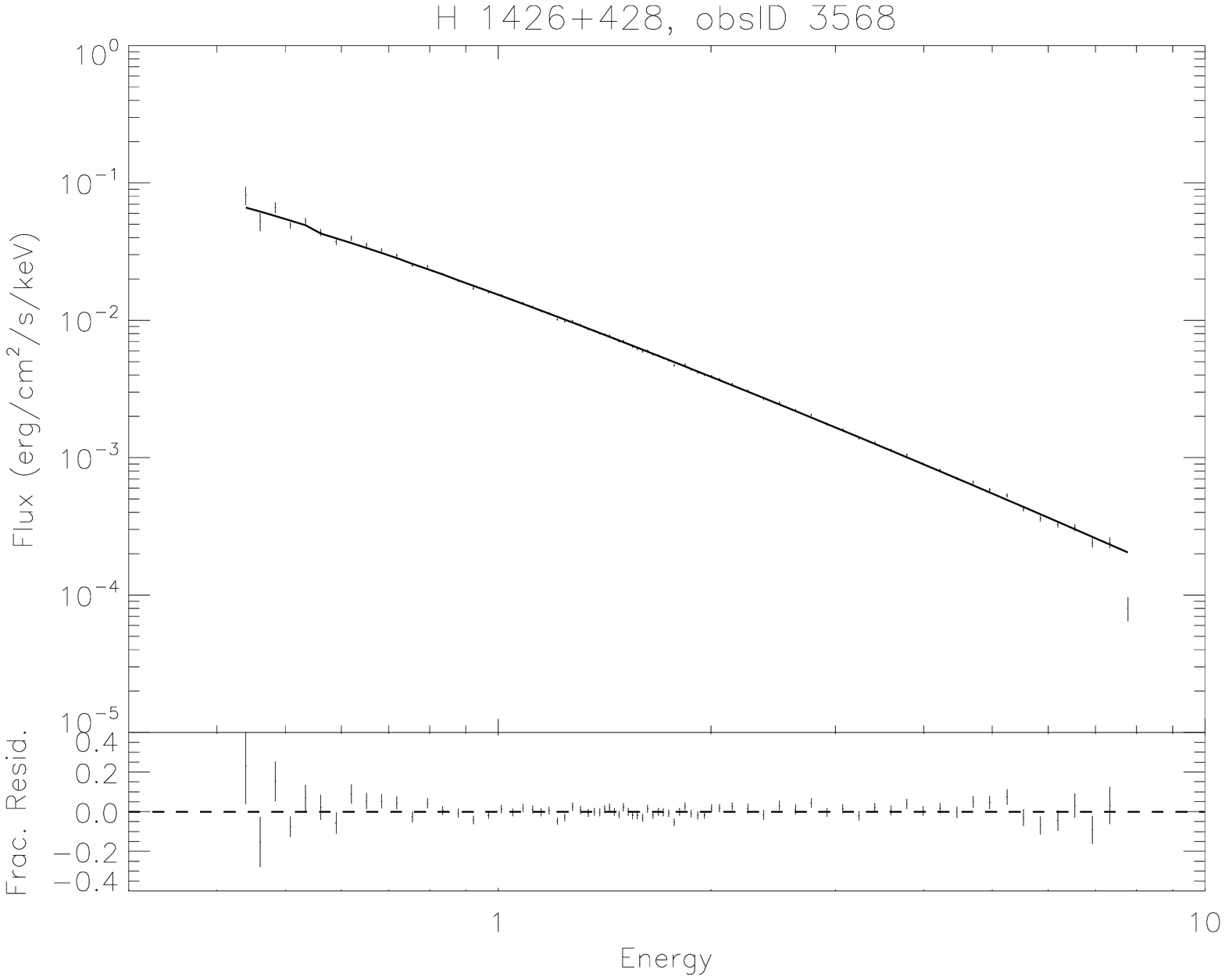}
   \end{tabular}
   \end{center}
\caption{
 Examples of spectral fits using eq.~\ref{eq:model} for two sources, the radio loud quasar 3C 273
 (obsID 3456, top) and the BL Lac object 1H 1426$+$428 (obsID 3568, bottom).
 The top panel shows the flux density and the fit and the bottom panel shows the residuals.
 Both are very well modeled by a power law with a slight curvature.
 In the case of 3C 273, there is positive curvature due to the soft excess and a
 possible reflection component or broadened Fe-K line.
 In this case, the reduced $\chi^2$ is over 2.
 For 1H 1426$+$428, the model fits much better and curves downward to high energies.
 }
\label{fig:fits}
\end{figure}

The data for the sources in Table~\ref{tab:observations} were fit to several
models to determine which sources have the smoothest spectra as
a function of $\lambda_x$.
Simple power law fits were tried first.
All were found to have spectra approximating a simple power law but
most showed significant spectral curvature.  The BLL sources
steepened to high energy while the AGN types generally flattened.
This curvature can be described using a simple logarithmic parabola
model with four parameters:

\begin{equation}
n_E = A e^{-N_H \sigma(E)} E^{-\Gamma + \beta \log E}
\label{eq:model}
\end{equation}

\noindent
where $N_H$ was fixed for each AGN to an estimate based on 21 cm
data (the results are robust against the exact choice), and the overall
model is  a power law with a curvature term, $\beta$.
The model is the same as that used for BL Lac objects by
Perlman et al.\ (2005 \cite{perlman05}).  The curvature term
is negative for spectra that are convex upward and
positive for those with soft or hard excesses.
Fig.~\ref{fig:fits} shows two examples of fits to sources in the list.

\begin{figure}
   \begin{center}
   \begin{tabular}{c}
   \includegraphics[width=15cm]{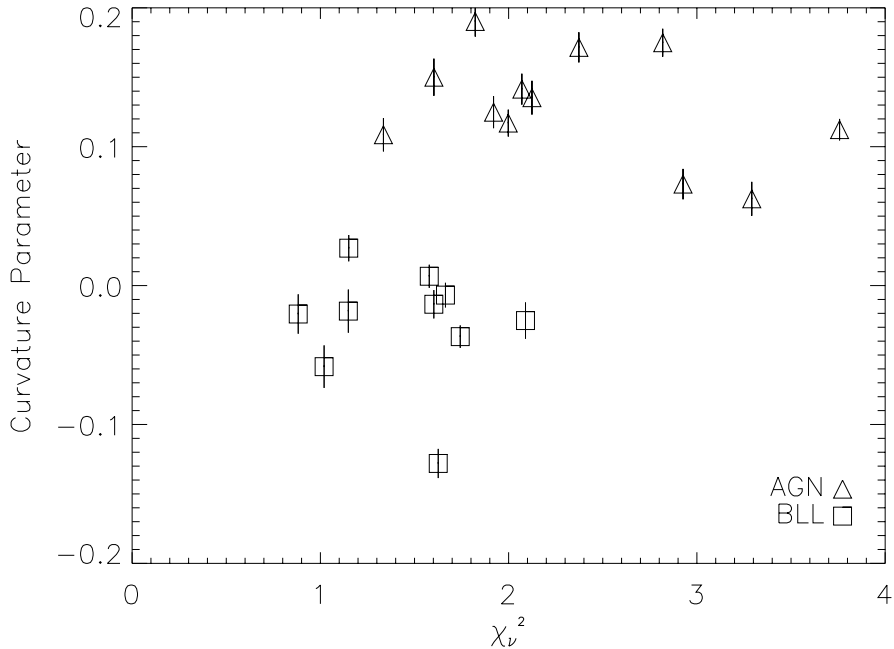}
   \end{tabular}
   \end{center}
\caption{
 The curvature parameter $\beta$, as a function of reduced $\chi^2$ for fits to
 HETGS spectra from Table~\ref{tab:observations}.
 See Eq.~\ref{eq:model} for the model, which defines $\beta$, the curvature parameter.
 The BLL type sources fit the model best and generally have negative curvatures,
 so that their spectra steepen with energy.  The AGN, such as 3C 273, have more
 complex spectra that include a soft excess.  For these fits, the cross-over
 wavelength was set to 10.3\AA.
}
\label{fig:chisq}
\end{figure}

Fig.~\ref{fig:chisq} shows that the quality of the fit is related to the
value of $\beta$: better fits are obtained for negative curvature.
The fits with positive curvature were predominantly radio-loud AGN
with optical emission lines, indicating that these spectra are actually more
complex than a simply curving model provides.  For the remaining
analysis, the sources with $\beta < 0.3$ were chosen; these were
the BL Lac objects PKS 2155-304 and 1H 1426-428.
The curvature values are similar to those found for other BL Lac objects
\cite{perlman05}.

\section{Determining the Cross-over Wavelength}

\begin{figure}
   \begin{center}
   \begin{tabular}{c}
   \includegraphics[width=15cm]{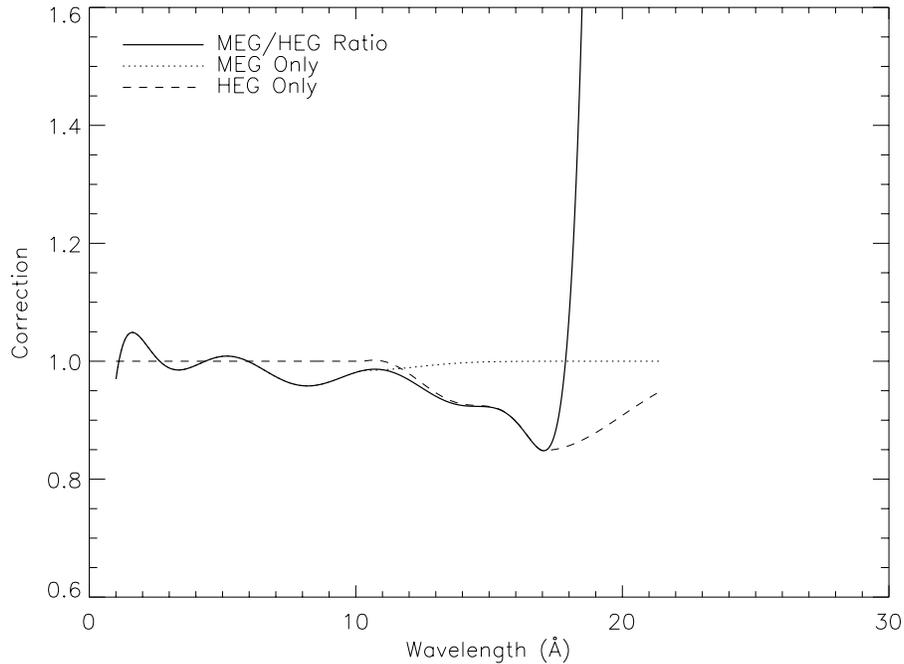}
   \end{tabular}
   \end{center}
\caption{
 Apportioning the MEG/HEG ratio correction.
 The correction was divided into two parts.
 Below 12 \AA, it is allocated to the MEG, while above 12 \AA, it is allocated primarily to the MEG. 
 The correction to the HEG efficiency is actually the inverse of that shown, so that it
 matches the expected total correction near 15 \AA.
 The HEG correction tails off to unity longward
 of 17 \AA\ due to the lack of data that causes the spline fit to extrapolate badly.
}
\label{fig:meghegfix}
\end{figure}

By fitting simple power law spectra to the BLL sources in the list,
one may argue against the extreme values of $\lambda_x$.
The average reduced
$\chi^2$, $\chi_\nu^2$, for the BLL
sources was smallest for $\lambda_x = 10.8$\AA, at 1.40;
for $\lambda_x = 5.3$\AA, the average $\chi_\nu^2$ was 2.78, and
for $\lambda_x = 17.7$\AA, the average $\chi_\nu^2$ was 1.68.
Thus, choosing a cross-over near 1 keV provides the smoothest
spectra.  A schematic of how the MEG/HEG ratio correction is
allocated is shown in Fig.~\ref{fig:meghegfix}.

The spectral residuals for the BL Lac objects were combined in order
to see if there were significant remaining residuals that could be
corrected.
Fig.~\ref{fig:resid} shows the result.  At this point, the residuals are
in the range of $\pm 3-5$\% over most of the HETGS energy
range and show no particular pattern that
could be readily corrected.

\section{Conclusion}

\begin{figure}
   \begin{center}
   \begin{tabular}{c}
   \includegraphics[width=18cm]{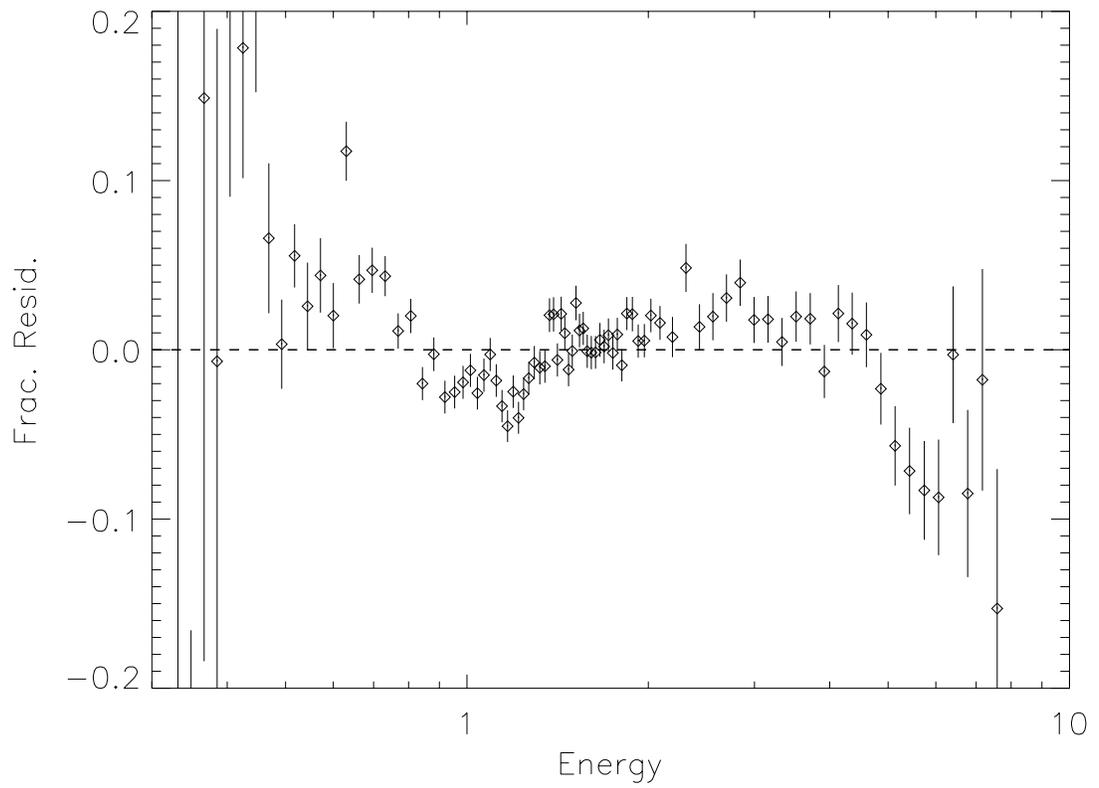}
   \end{tabular}
   \end{center}
\caption{
 The average residuals for curved power law fits to the HETGS
 data for BL Lac objects.  For most of the HETGS range, the
 systematic deviations are not significant or are less than 3\%.
 Deviations of 5-10\% near 6 keV and 0.7 keV will be investigated
 further.
}
\label{fig:resid}
\end{figure}

While there are possible systematic errors of up to 5\% over
the 0.5-7 keV range, most of the deviations are less than 3\%
(Fig.~\ref{fig:resid}).
Residuals to simple fits to the other AGN (not shown) are also less than 3\%
over most of the 0.5-7\% range and only as large as 5\% near 0.7 keV.
Thus, an approximate limit to relative systematic errors is about
3-5\% over the HETGS range.
While these residual errors will be examined further to see if they
can be eliminated, another area of investigation will be the cross-dispersion
selection efficiency, which is currently applied in the grating RMFs.

\acknowledgments     

Support for this work was provided by the National Aeronautics and Space Administration
(NASA) through the Smithsonian Astrophysical Observatory (SAO) contract SV3-73016
to MIT for support of the Chandra X-Ray Center (CXC), which is operated by SAO for
and on behalf of NASA under contract NAS8-03060.


\begin{thebibliography}{}
\bibitem[Canizares et al.(2005)]{hetg} Canizares, C.R., et al., 2005, PASP,
	117, 1144.
\bibitem[Marshall et al.(2004)]{2004SPIE.5165..457M} Marshall, H.~L., 
	Dewey, D., \& Ishibashi, K.\ 2004, \procspie, 5165, 457 
\bibitem[Marshall et al.(1998)]{1998SPIE.3444...64M} Marshall, H.~L., 
	Dewey, D., Schulz, N.~S., \& Flanagan, K.~A.\ 1998, \procspie, 3444, 64 
\bibitem[Perlman et al.(2005)]{perlman05} Perlman, E.~S., et al.\ 
	2005, \apj, 625, 727 

\end{thebibliography}
\end{document}